\pdfoutput=1

\documentclass[11pt]{article}
\usepackage{amssymb}
\usepackage{acl}

\usepackage{times}
\usepackage{latexsym}

\usepackage[T1]{fontenc}

\usepackage[utf8]{inputenc}

\usepackage{microtype}

\usepackage{inconsolata}

\usepackage{soul}
\usepackage{graphicx}
\usepackage{booktabs}
\usepackage{float}
\usepackage{overpic}
\usepackage{float}  
\usepackage{subfloat}
\usepackage{stfloats} 
\usepackage[skip=0.5\baselineskip]{caption}
\usepackage{bm}
\usepackage{amsmath}
\usepackage{booktabs}
\usepackage{multirow}
\usepackage{multicol}
\usepackage{enumitem}
\usepackage{subcaption}
\usepackage{threeparttable}
\usepackage{xspace}
\usepackage{color}
\usepackage[normalem]{ulem}
\usepackage{algorithmicx,algorithm}
\usepackage{algpseudocode}
\usepackage{algorithm,algpseudocode}
\usepackage{marvosym}   
\usepackage{colortbl}   
\usepackage{tcolorbox}  
\tcbuselibrary{breakable}   
\usepackage{wasysym}    
\usepackage[figuresright]{rotating} 

\usepackage{xspace}

\newcommand{\eg}{\emph{e.g.,}\xspace}
\newcommand{\ie}{\emph{i.e.,}\xspace}

\newcommand{\methodname}{D$^2$MDT\xspace}

\title{\methodname: Department-aware Multidisciplinary Team Consultation with Deliberation for Efficient Clinical Prediction}

\author{
Yongqi Liang,
Qidong Liu\textsuperscript{*},
Chunze Yang,
Lei Wu,
Jiusong Ge,
Ni Zhang,
Chen Li\textsuperscript{*} \\
Xi'an Jiaotong University, Xi'an, China \\
\texttt{\{LYQi, pureeeee, 2216113083, jiusongge\}@stu.xjtu.edu.cn} \\
\texttt{\{liuqidong, nizhang, cli\}@xjtu.edu.cn}\\
\textsuperscript{*}Corresponding authors.
}

\begin{document}

\maketitle

\begin{abstract}
Electronic health records (EHRs) are central to clinical prediction, but existing methods either rely on correlation-driven deep models or use single large language models (LLMs), making it difficult to support multidisciplinary clinical reasoning. Recent multi-agent systems (MAS) provide a promising alternative, yet current EHR-grounded MAS methods still suffer from weak evidence differentiation across agents and redundant multi-round interaction. We propose \textbf{\methodname}, 
a \textbf{D}epartment-aware \textbf{M}ulti\textbf{D}isciplinary \textbf{T}eam Consultation with \textbf{D}eliberation
for \textbf{Efficient} clinical prediction. 
\methodname first constructs structured EHR evidence and consultation-ready semantic evidence for multi-agent consultation. It then assigns patient-specific department perspectives to doctor agents and retrieves complementary evidence for collaborative consultation. To improve efficiency, \methodname further introduces residual deliberation, which updates only unresolved consensus rather than replaying the full discussion history. Finally, \methodname fuses the refined consensus report with structured EHR representations for prediction. Experiments on mortality prediction show that \methodname improves both predictive performance and consultation efficiency.
We release the code online to ease the reproducibility of this paper\footnote{https://github.com/GigiResearch/D2MDT}.
\end{abstract}

\section{Introduction}

Electronic health records (EHRs) have become a central foundation for data-driven clinical decision support, where patient trajectories are modeled to anticipate future risks~\cite{2025TrajSurv} and support timely intervention~\cite{2025Early}. 
Existing deep learning methods generally learn predictive patterns directly from EHR sequences, such as AdaCare~\cite{2020AdaCare} and PAI~\cite{2024Learnable}.
However, most deep EHR models remain primarily correlation-driven, making it difficult to explicitly capture clinical semantics and incorporate medical knowledge. 
This limitation has motivated growing interest in Large Language Models (LLMs)-based EHR modeling,
where language models can connect structured patient records with semantic reasoning and natural-language interpretation~\cite{2025A,du2026testing}.

While LLMs offer potential for semantic EHR modeling, complex clinical decision-making is inherently collaborative rather than individual~\cite{2024Collaborative,2024Evaluation}. 
In general, difficult cases require multidisciplinary team (MDT) consultation, where experts with complementary knowledge jointly examine clinical evidence, compare diagnostic hypotheses, and refine treatment decisions through structured discussion~\cite{2015Improving,2025Hospital,graber2017new,2022Factors}. 
This indicates the necessity of MDT in the EHR prediction task.
However, MDT consultation is difficult to replicate by single LLMs.
Despite LLMs' ability to understand cases by clinical semantics,
it still tends to compress the case into one dominant reasoning trajectory, making it difficult to preserve parallel specialty-specific assessments or coordinate iterative cross-perspective discussion~\cite{2024Evaluation,2024Accuracy}. 

Recently, Multi-Agent Systems (MAS) have emerged as a promising approach for simulating MDT-style clinical reasoning in healthcare by incorporating diverse clinical perspectives and interpretable intermediate discussions. 
Earlier research studies~\cite{2024MDAgents,2024Beyond,2025Sequential,2026EvoMDT}, denoted as \textbf{Linguistic MAS}, 
show that multi-agent deliberation can improve semantic clinical reasoning. However, these frameworks operate primarily over textual case understanding, so their reasoning remains only weakly tied to structured patient measurements and is therefore limited for rigorous EHR-based prediction.
To address this issue, \textbf{EHR-aware MAS} works,
such as 
ColaCare~\cite{Wang2025ColaCare}, 
and ClinNoteAgents~\cite{2025ClinNoteAgents} strengthen EHR grounding and bring collaborative reasoning closer to quantitative patient-level prediction. 
\begin{figure}[!t]
\centering
\includegraphics[width=1\linewidth]{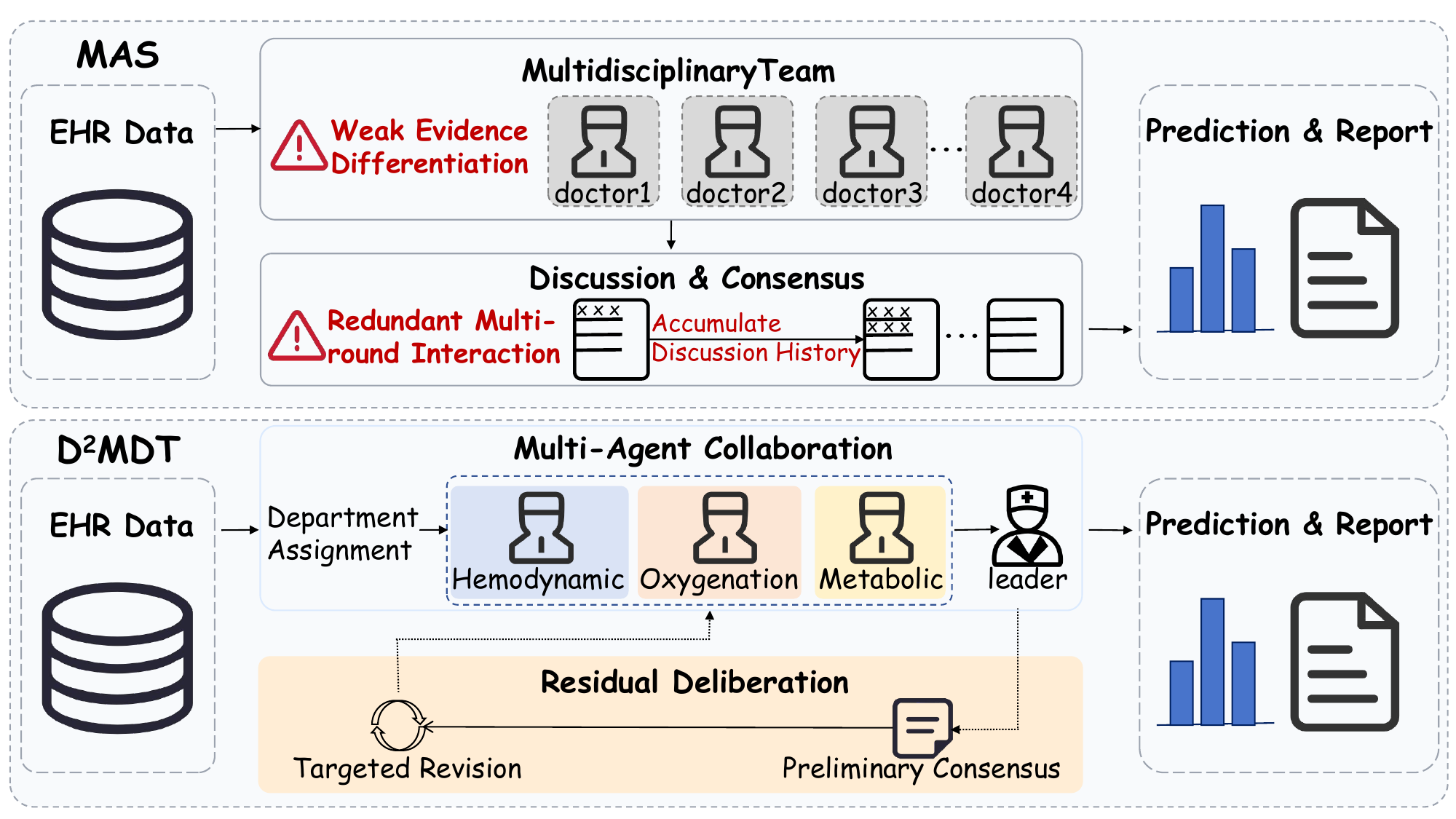}
\caption{The illustration of \methodname and existing MAS.}
\label{fig:preliminary}
\vspace{-5mm}
\end{figure}

However, there are still two challenges faced by EHR-aware MAS methods, as illustrated in Figure~\ref{fig:preliminary}.
\textbf{i) Weak Evidence Differentiation}. Multi-agent frameworks can improve apparent diversity by assigning different roles to different agents~\cite{2024EHRAgent,2024MDAgents,2024MedAgents,2025MedAgentBench,2025MedAgentBoard}, but these roles are usually only weakly grounded in patient-specific structured evidence, and therefore role diversity does not necessarily translate into evidence diversity. \textbf{ii) Redundant Multi-round Interaction}. As consultation rounds accumulate, later interactions repeatedly revisit resolved content, leading to substantial context redundancy and reduced reasoning efficiency~\cite{Wang2025ColaCare,2025MDTeamGPT}. These two issues jointly limit both the reliability and the scalability of multi-round clinical consultation for EHR prediction.

To address the challenges mentioned above, we propose \methodname, a Department-aware MultiDisciplinary Team Consultation with Deliberation for Efficient clinical prediction. 
First, \methodname introduces Department Agents to handle the issue of weak evidence differentiation.
Instead of relying on loosely specified agent roles, \methodname assigns patient-specific department perspectives to doctor agents and couples them with department-aware evidence retrieval.
Then, \methodname reduces redundant multi-round interaction by carrying forward only the unresolved part of the current consensus, allowing later rounds to focus on disagreement refinement rather than repeated discussion of the full patient context.
The contributions are as follows:
\begin{itemize}[itemsep=0pt, topsep=1pt, leftmargin=*]
    \item Insightfully, we introduce a department-aware collaborative agent framework that explicitly simulates department-level consultation for EHR-based clinical prediction. 
    \item Technically, we design a fusion mechanism that integrates Department Agent reasoning with model-agent predictions, enabling structured coordination between numerical EHR models and language-based clinical reasoning.
    \item 
    We empirically evaluate \methodname on two EHR datasets and show that it achieves better predictive performance with higher consultation efficiency in multi-round settings.
\end{itemize}


\section{Preliminary}

In this section, the problem definition of the prediction task is given. Specifically, we focus on predicting patient mortality outcomes. Our goal is to extract knowledge from EHR data, supplemented by external medical knowledge (such as clinical guidelines), to enhance predictive modeling of electronic health records. Therefore, the prediction target is expressed as:
\begin{equation}
    \hat{y} = G(\mathbf{x}_{\text{EHR}}, \text{MedicalKnowledge}) 
\end{equation}
where $\hat{y}$ is the prediction outcome.
For example, $\hat{y}=0$ indicates the patient is alive in the mortality prediction task.
$G(\cdot,\cdot)$ is the department framework. 
$\mathbf{x}_{\text{EHR}}$ is the structure of the EHR dataset which is multivariate time series data with multiple features, represented as 
$\mathbf{X}=[\mathbf{x_{1}},\mathbf{x_{2}},\ldots,\mathbf{x_{t}}]^{\top}\in\mathbb{R}^{T\times F}$, encompassing information across $T$ visits and $F$ features, including 
dynamic features (\eg laboratory tests and vital signs). 


\section{Method}
\begin{figure*}[t]
\centering
\includegraphics[width=0.9\linewidth]{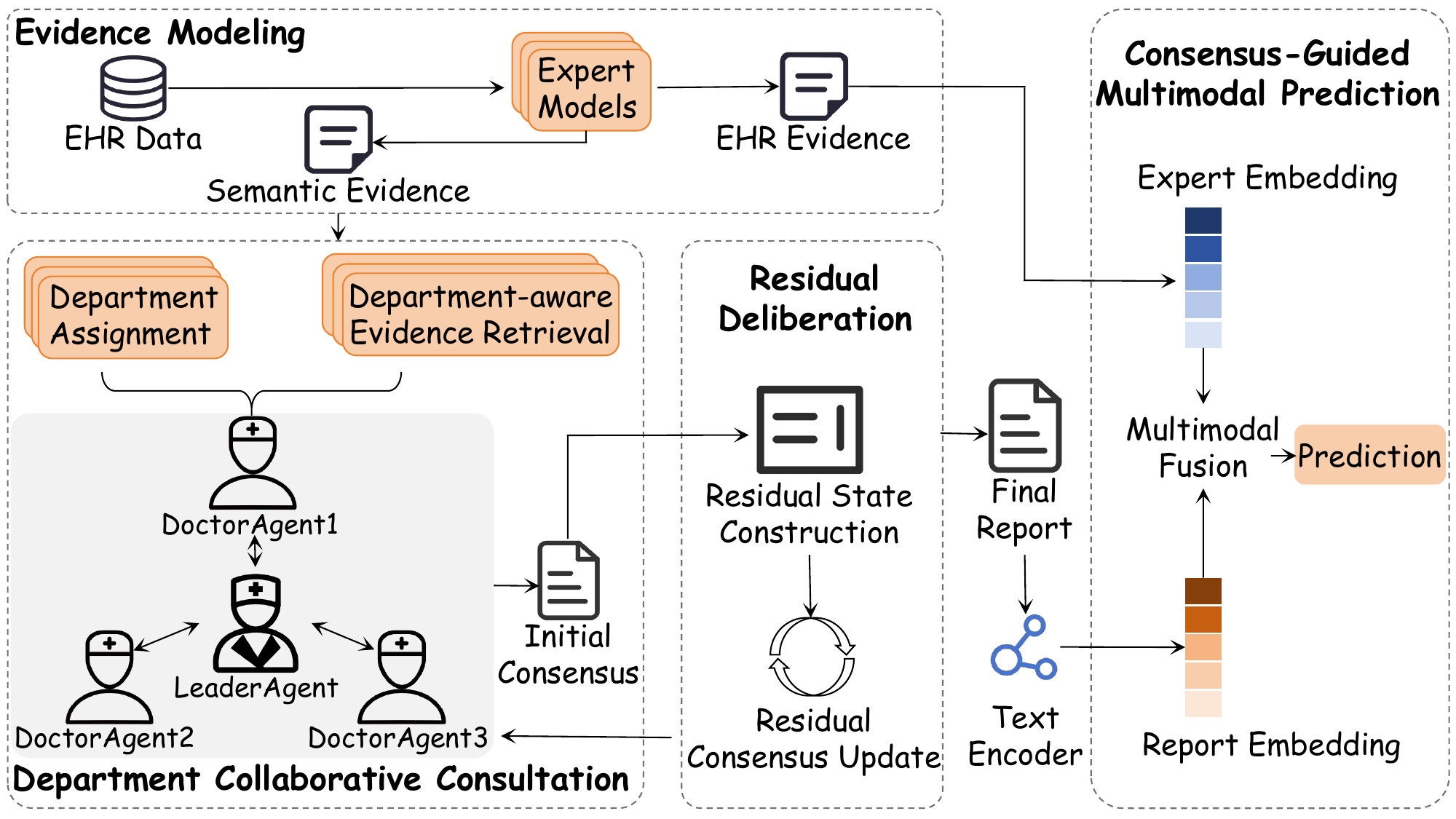}
\caption{The overview of the proposed \methodname.}
\label{fig:framework}
\vspace{-2mm}
\end{figure*}
\subsection{Overview}
We propose \textbf{\methodname}, a Department-aware Multidisciplinary Team Consultation with Deliberation for Efficient clinical prediction. As illustrated in Figure~\ref{fig:framework}, to address the weak evidence differentiation and redundant multi-round interaction, \methodname first performs \textbf{Evidence Modeling} to derive structured EHR evidence and consultation-ready semantic evidence. However, evidence views alone do not guarantee clinically meaningful specialist diversity. \methodname therefore conducts \textbf{Department Collaborative Consultation}, where doctor agents are assigned patient-specific department perspectives and access complementary external evidence to form an initial patient-level consensus. On top of this, \methodname applies \textbf{Residual Deliberation} to carry forward only the unresolved part of the current consensus, so that later rounds refine disagreement through compact updates instead of replaying the entire interaction history. Finally, \methodname performs \textbf{Consensus-Guided Multimodal Prediction} by fusing the refined consensus report with structured EHR representations for final risk estimation. In this way, \methodname turns collaboration into an evidence-grounded and refined reasoning process for clinical prediction.

\subsection{Evidence Modeling}
This section aims to build evidence of \methodname. Since consultation should be grounded in patient-specific clinical signals rather than latent vectors alone, we construct both structured \textbf{EHR Evidence} for prediction and \textbf{Semantic Evidence} for Multi-Agent consultation from records.

\vspace{1mm}
\noindent\textbf{EHR Evidence.}
For a patient, let $b$ denote the basic profile, including static demographics and admission-related information, 
Let $\{E_m\}_{m=1}^{M}$ denote the set of pretrained EHR expert models, where the $m$-th expert $E_m$ contains a temporal encoder and its prediction head. \methodname applies all experts to the same patient trajectory and obtains:
\begin{equation}
\mathbf{h}_m,\hat{y}_m = E_m(\mathbf{X}) \qquad m=1,\ldots,M,
\end{equation}
where $\mathbf{h}_m\in\mathbb{R}^{d_m}$ is the structured patient representation produced by expert $E_m$, and $\hat{y}_m\in(0,1)$ is its preliminary risk estimate. These expert-specific hidden representations preserve complementary temporal patterns and are later used in multimodal fusion. During fusion training, the pretrained expert models are kept fixed.

\vspace{1mm}
\noindent\textbf{Semantic Evidence}.
Numerical embeddings alone are insufficient for language-based consultation. 
\methodname therefore converts each expert output into an interpretable evidence card in two steps. For expert $E_m$, we first compute feature attributions by $\boldsymbol{\alpha}_m = \operatorname{SHAP}(E_m, \mathbf{X})$~\cite{sundararajan2020many}, where SHAP estimates feature attribution scores based on Shapley values to quantify each variable's contribution to the expert model's risk estimate. We retain the top-$K_m$ important variables $f_{m,k}$. We then verbalize each retained variable with its name, latest value, and short local trend near time $T$ (e.g., persistent elevation, recent drop, or abnormal fluctuation). We denote the resulting expert-specific evidence card by:
\begin{equation}
\mathcal{E}_m = \{(f_{m,k}, u_{m,k}, \alpha_{m,k})\}_{k=1}^{K_m}
\end{equation}
where $u_{m,k}$ is the textual status summary, and $\alpha_{m,k}$ is the attribution score measuring how strongly this variable supports the expert prediction. \methodname then forms the expert consultation context:
\begin{equation}
c_m = \big[b;\operatorname{Recent}(\mathbf{X});\hat{y}_m;\mathcal{E}_m\big]
\end{equation}
where $\operatorname{Recent}(\cdot)$ denotes a short summary of the most recent observations and local temporal changes. The resulting context summarizes the patient profile, expert-estimated risk, and salient abnormalities to expert $E_m$. For department routing, \methodname further uses the patient-level summary $c^{\mathrm{route}}=[b;\operatorname{Recent}(\mathbf{X});\{(\hat{y}_m,\mathcal{E}_m)\}_{m=1}^{M}]$, which aggregates all expert-specific evidence cards into a unified routing view. The retained fields are provided in Appendix~\ref{sec:appendix-semantic} for implementation clarity.

\subsection{Department Collaborative Consultation}
Building on the evidence views above, \methodname organizes multi-agent consultation into a department-aware specialist collaboration process. This process consists of three consecutive stages: \textbf{Department Assignment}, which allocates 
specialist identities to doctor agents; \textbf{Department-aware Evidence Retrieval}, which enables each specialist to access complementary external evidence under its assigned department perspective; and \textbf{Initial Consensus}, which aggregates the resulting specialist opinions into a patient-level preliminary report.

\vspace{1mm}
\noindent\textbf{Department Assignment}.
\methodname constructs three routing cues. First, it extracts the set of abnormal measurements $\mathcal{U}^{\mathrm{abn}}$ from the patient trajectory by flagging variables whose latest value deviates markedly from that patient's recent range.
Second, it aggregates the expert-level important feature sets $\mathcal{U}^{\mathrm{imp}}=\bigcup_{m=1}^{M}\{f_{m,k}\}$ into a patient-level important feature set, $f_{m,k}$ is the important feature set of expert $E_m$, 
Third, it extracts a context-based risk signal $c_{d_i}$ from patient-level summary $c^{\mathrm{route}}$, which preserves coarse global cues such as predicted mortality risk, ICU status, and high-risk percentile information.
For each candidate department $d_i\in\mathcal{D}$, \methodname computes a routing score:
\begin{equation}
\begin{aligned}
s_{d_i}
&=
\sum_{f\in\mathcal{U}^{\mathrm{abn}}} 
\lambda_{\mathrm{abn}}\, \rho_{d_i}(f)
\\
&+
\sum_{f\in\mathcal{U}^{\mathrm{imp}}} 
\lambda_{\mathrm{imp}}\, \rho_{d_i}(f)
+
\lambda_{\mathrm{ctx}}\, c_{d_i}
\end{aligned}
\end{equation}
where $\rho_{d_i}(\cdot)$ denotes the match score between variable $f$ and department $d_i$.
Therefore, the three terms correspond to abnormal-numeric evidence, important-feature evidence, and global context-risk evidence, respectively.  $\lambda_{\mathrm{abn}}$, $\lambda_{\mathrm{imp}}$, and $\lambda_{\mathrm{ctx}}$ are their balancing weights. The assigned departments are then selected by ranking these scores:
\begin{equation}
(d_1,\ldots,d_M)=\operatorname{Top_{M}}\big(\{s_d\mid {d_i}\in\mathcal{D}\}\big)
\end{equation}
In implementation, one hemodynamic role is kept as a global anchor, while the other doctor agents are filled by the highest-scoring department candidates. More details are given in Appendix~\ref{sec:appendix-department}.

\vspace{1mm}
\noindent\textbf{Department-aware Evidence Retrieval}.
After department assignment, each doctor agent reasons over the same patient case through a department-specific clinical lens. For doctor agent $D_m$, \methodname constructs a department-aware consultation context by adding a department hint, and constructs the retrieval query $q_m$ by appending department keywords to the patient summary. In this way, the same patient case is viewed under different specialist emphases. The exact prompt and templates are listed in Appendix~\ref{sec:appendix-prompt-template}. The doctor then retrieves complementary medical evidence by: 
\begin{equation}
z_m=\operatorname{Retrieve}(q_m)
\end{equation}
where $q_m$ denotes the resulting department-aware query and $z_m$ denotes the corresponding retrieved evidence. This design ensures that different doctor agents differ not only in role identity but also in the supporting evidence they consult.

\vspace{1mm}
\noindent\textbf{Initial Consensus}.
Given the department-aware patient view and the corresponding retrieved external evidence $z_m$, the $m$-th doctor agent produces a structured specialist opinion:
\begin{equation}
o_m^{(0)}=\operatorname{Doctor}_m(b,z_m)
\end{equation}
where $o_m^{(0)}$ contains the doctor's risk judgment, supporting rationale, and cited evidence grounded in the department-specific patient view. The leader agent then aggregates all specialist opinions into a preliminary patient-level report:
\begin{equation}
R^{(0)}=\operatorname{Leader}\big(b,\{o_m^{(0)}\}_{m=1}^{M}\big)
\end{equation}
This report serves as the initial consensus for the subsequent residual deliberation process.

\subsection{Residual Deliberation}
Although department collaborative consultation improves evidence diversity, later-round interaction can still become inefficient if all agents continue exchanging full narrative histories. \methodname therefore proposes a residual deliberation process, in which each round focuses only on the unresolved part of the current consensus. Concretely, this process involves two steps: \textbf{Residual State Construction}, which separates stable consensus from remaining disagreement, and \textbf{Residual Consensus Update}, 
which guides doctor agents to revise their opinions based on the residual disagreement.

\vspace{1mm}
\noindent\textbf{Residual State Construction.} After round $r$, the leader report $R^{(r)}$ is compressed into a settled consensus state $R_C^{(r)}$ and a residual disagreement state $R_{\Delta}^{(r)}$, $R_C^{(r)}$ stores stable fields such as current risk level, consensus strength, and agreed mechanisms, while $R_{\Delta}^{(r)}$ keeps only unresolved discussion points such as minority view, risk disagreement, mechanism mismatch, or escalation cues. 
Each specialist opinion is compressed into a compact doctor state:
\begin{equation}
\begin{aligned}
s_m^{(r)}
=
\big[
&v_m^{(r)};\ \mu_m^{(r)};\ \psi_m^{(r)};
\\
&\eta_m^{(r)};\ \gamma_m^{(r)}
\big]
\qquad m=1,\ldots,M .
\end{aligned}
\end{equation}
where $v_m^{(r)}$ is the doctor's risk vote, $\mu_m^{(r)}$ summarizes top mechanisms, $\psi_m^{(r)}$ records key support features, $\eta_m^{(r)}$ records counter-evidence, and $\gamma_m^{(r)}$ represents confidence and escalation status. 
As a result, later rounds operate on compact structured summaries rather than full narrative histories.

\vspace{1mm}
\noindent\textbf{Residual Consensus Update.}  At round $r+1$, doctor agent $D_m$ updates its opinion by reusing its department-aware patient view and retrieved external evidence while receiving only the compact residual messages from the previous round:
\begin{equation}
o_m^{(r+1)}=\operatorname{Doctor}_m\big(\,s_m^{(r)},R_{\Delta}^{(r)},R_C^{(r)}\big)
\end{equation}
Here 
$s_m^{(r)}$, $R_{\Delta}^{(r)}$, and $R_C^{(r)}$ are compact residual states transmitted between rounds. Thus, later rounds do not replay earlier detailed opinions; they only pass the concise doctor card and compact leader report. The leader then revises the global report by integrating the specialist updates with the current consensus and residual disagreement state:
\begin{equation}
R^{(r+1)}=\operatorname{Leader}(R_C^{(r)},R_{\Delta}^{(r)},\{o_m^{(r+1)}\}_{m=1}^{M}\big)
\end{equation}
Therefore, \methodname performs multi-round consultation as a sequence of residual corrections, where each round only refines what remains unresolved. The discussion terminates when the residual state becomes empty or sufficiently weak, or when the maximum number of rounds is reached. Let $K$ denote the total number of executed rounds and let $R^{\ast}=R^{(K)}$
be the final consensus report.
\subsection{Consensus-Guided Multimodal Prediction}
The final consensus report is not used only for interpretability, it is also treated as an additional predictive modality. \methodname first encodes the final consultation report by a clinical text encoder:
\begin{equation}
\mathbf{h}_{\mathrm{txt}}=E_{\mathrm{txt}}(R^{\ast})
\end{equation}
where $E_{\mathrm{txt}}(\cdot)$ denotes the report encoder and $\mathbf{h}_{\mathrm{txt}}$ is the resulting text representation. \methodname then fuses this representation with the structured EHR representations from all expert models:
\begin{equation}
\mathbf{h}_{\mathrm{fus}}=
\phi\!\left([\mathbf{h}_1;\cdots;\mathbf{h}_M;\mathbf{h}_{\mathrm{txt}}]\right)
\end{equation}
where $[\cdot;\cdot]$ denotes vector concatenation and $\phi(\cdot)$ is a 
fusion network. The final prediction is as follows:
\begin{equation}
\hat{y}=\sigma(\mathbf{w}^{\top}\mathbf{h}_{\mathrm{fus}}+b_0)
\end{equation}
where $\mathbf{w}$ and $b_0$ are the predictor parameters, and $\sigma(\cdot)$ is the sigmoid function. For a training set with $N$ patients, the model is optimized by the binary cross-entropy objective:
\begin{equation}
\mathcal{L}_{\mathrm{pred}}
=
-\frac{1}{N}
\sum_{i=1}^{N}
\left(
y_i\log\hat{y}_i
+
(1-y_i)\log(1-\hat{y}_i)
\right)
\label{eq:loss}
\end{equation}
where $y_i\in\{0,1\}$ is the ground-truth clinical outcome of patient $i$, and $\hat{y}_i$ is the predicted risk. This design allows \methodname to combine fine-grained temporal EHR dynamics with high-level consensus reasoning formed through department-aware and residual-efficient consultation.

\subsection{Training and Inference}
\noindent\textbf{Training}.
Each specialist EHR encoder is trained independently to extract patient-specific hidden representations for all samples.
The fusion predictor then takes the specialist EHR representations together with $\mathbf{h_{\text{txt}}}$ as input and is optimized with the binary cross-entropy loss in Eq.~\eqref{eq:loss}. The pretrained specialist EHR encoders are kept fixed.

\vspace{1mm}
\noindent\textbf{Inference}.
Given a patient’s EHR sequence, the pretrained specialist encoders first produce the hidden representations $\mathbf{h}_m$. The department collaborative consultation module then generates the final consensus report through multi-round reasoning, and the report is encoded as $\mathbf{h_{\text{txt}}}$. Finally, the fusion module combines $\mathbf{h}_m$ and $\mathbf{h_{\text{txt}}}$ to produce the mortality prediction $\hat{y}$. Due to space limitation, we leave the algorithm to Appendix~\ref{sec:appendix:method_algorithm}.

\section{Experiment}

\subsection{Experimental Setup}
\begin{table*}[!t]
\centering
\caption{The overall results of competing baselines and \methodname on MIMIC-III Outcome and MIMIC-IV Outcome. The \textbf{boldface} refers to the highest score and the \underline{underline} indicates the second-best result.}
\label{tab:mimic_results}
\resizebox{\textwidth}{!}{
\begin{tabular}{l|ccc|ccc}
\toprule
\multirow{2}{*}{\textbf{Methods}} 
& \multicolumn{3}{c|}{\textbf{MIMIC-III Outcome}} 
& \multicolumn{3}{c}{\textbf{MIMIC-IV Outcome}} \\
& AUPRC ($\uparrow$) & AUROC ($\uparrow$) & min(+P, Se) ($\uparrow$)
& AUPRC ($\uparrow$) & AUROC ($\uparrow$) & min(+P, Se) ($\uparrow$) \\
\midrule
AdaCare & 48.24$\pm$3.23 & 80.30$\pm$1.70 & 46.45$\pm$3.38 & 60.51$\pm$4.47 & 85.85$\pm$2.14 & 57.35$\pm$3.90 \\
RETAIN & 50.51$\pm$4.42 & \underline{84.23$\pm$1.73} & \underline{50.88$\pm$3.74} & 57.43$\pm$5.55 & 87.10$\pm$2.14 & 58.04$\pm$4.41 \\
PAI & 49.72$\pm$4.11 & 81.76$\pm$1.91 & 49.77$\pm$3.23 & 54.68$\pm$5.56 & 85.07$\pm$2.35 & 57.65$\pm$4.47 \\

\midrule
EMERGE & \underline{50.82$\pm$3.74} & 83.49$\pm$1.72 & 50.35$\pm$2.62 & 58.74$\pm$4.39 & \underline{87.85$\pm$1.68} & 57.27$\pm$3.38 \\
MedGemma$_{\mathrm{ZeroShot}}$ & 28.64$\pm$3.21 & 69.11$\pm$2.58 & 31.16$\pm$2.87 & 28.61$\pm$3.74 & 73.31$\pm$2.85 & 34.52$\pm$3.92 \\
MedGemma$_{\mathrm{FewShot}}$ & 14.97$\pm$1.24 & 56.28$\pm$1.42 & 15.18$\pm$1.26 & 11.79$\pm$1.15 & 39.10$\pm$1.53 & 12.43$\pm$1.21 \\
\midrule
MDAgents & 45.72$\pm$4.44 & 82.31$\pm$1.89 & 48.41$\pm$3.42 & 53.49$\pm$5.12 & 85.57$\pm$1.97 & 54.42$\pm$4.17 \\
MDTeamGPT & 48.31$\pm$3.32 & 80.19$\pm$1.88 & 44.62$\pm$2.93 & 56.87$\pm$4.8 & 81.47$\pm$2.6 & 53.05$\pm$4.2 \\
ColaCare & 50.81$\pm$5.05 & 84.07$\pm$2.01 & 50.52$\pm$4.19 & \underline{63.24$\pm$5.39} & \textbf{88.17$\pm$2.19} & \underline{61.03$\pm$4.52} \\
\midrule
\rowcolor{orange!10}\textbf{\methodname} & \textbf{53.72$\pm$4.67} & \textbf{84.56$\pm$2.07} & \textbf{54.68$\pm$3.70} & \textbf{65.31$\pm$4.41} & 87.39$\pm$1.97 & \textbf{63.66$\pm$3.91} \\
\bottomrule
\end{tabular}
}
\vspace{-2mm}
\end{table*}

\noindent\textbf{Datasets and Task.}
We evaluate \methodname on two widely used ICU benchmark datasets, MIMIC-III~\cite{johnson2016mimic,harutyunyan2019multitask} and MIMIC-IV~\cite{johnson2023mimic}. For both datasets, the task is in-hospital outcome prediction based on structured EHR records. Dataset statistics, splits, and details are deferred to Appendix~\ref{sec:appendix:datasets}.

\vspace{1mm}
\noindent\textbf{Baselines.}
We compare \methodname with three groups of baselines. The first group contains deep learning baselines, including AdaCare~\cite{2020AdaCare}, RETAIN~\cite{choi2016retain}, and PAI~\cite{2024Learnable}. The second group contains LLM-driven collaborative baselines, including EMERGE~\cite{zhu2024emerge} and MedGemma~\cite{sellergren2025medgemma}. The third group contains MDT baselines, including MDAgents~\cite{2024MDAgents} MDTeamGPT~\cite{2025MDTeamGPT} and ColaCare~\cite{Wang2025ColaCare}.   Appendix~\ref{sec:appendix:baseline} gives the details of baseline settings.

\vspace{1mm}
\noindent\textbf{Evaluation Metrics.}
We report \textbf{AUPRC}~\cite{kim2022empirical}, \textbf{AUROC}~\cite{mcdermott2024closer}, and \textbf{min(+P, Se)}~\cite{ma2022patient} for predictive performance. 
To evaluate efficiency, we additionally report Average Round-1 Prompt Token (\textbf{AR1PT}) and Average Round-2 Prompt Token (\textbf{AR2PT}). AR2PT being smaller than AR1PT indicates that residual discussion is effective during the collaboration process.
More details of evaluation metrics are given in the Appendix~\ref{sec:appendix:metrics}.

\vspace{1mm}
\noindent\textbf{Implementation Details.}
All consultation agents are driven by Qwen3-8B. For retrieval-augmented consultation, we use MedCPT 
with the MSD corpus. For multimodal prediction, the final leader report is encoded by GatorTron-base
and fused with the three expert EHR embeddings by a lightweight MLP with hidden dimension 128. 
Additional details are provided in the Appendix~\ref{sec:appendix:implementation}.

\subsection{Overall Performance}


We show the overall performance of our \methodname and competing baselines in Table~\ref{tab:mimic_results}. 
Overall, the proposed \methodname achieves the best performance on most key metrics across the two datasets, while remaining competitive on the remaining ones, demonstrating its overall effectiveness for EHR-based outcome prediction.
Then, a more detailed analysis of the results will be given.

Deep Learning baselines lag behind the stronger methods overall. This is mainly because 
they are limited in capturing complex clinical semantics and leveraging complementary knowledge from multiple medical perspectives. As a result, their performance improvements are constrained when facing challenging outcome prediction tasks.

LLM-based methods do not show clear advantages in this task. Although such methods introduce external medical knowledge or reasoning ability from LLMs, they still struggle to fully adapt to structured and temporal clinical prediction settings. 

MDT-based baselines perform better overall, which verifies the value of multi-disciplinary collaboration in clinical decision-making. By introducing interactions among different medical roles, these methods are able to provide more comprehensive evidence than conventional deep learning or vanilla LLM-based approaches. However, \methodname still shows stronger overall performance than these baselines, suggesting that our method can more effectively transform multi-disciplinary discussions into predictive signals. The advantage of \methodname indicates that structured collaboration and effective aggregation of expert opinions are crucial for improving clinical outcome prediction, especially under complex and high-risk medical scenarios.

\subsection{Efficiency Study}
\begin{figure}[!t]
\centering
\includegraphics[width=1\linewidth]{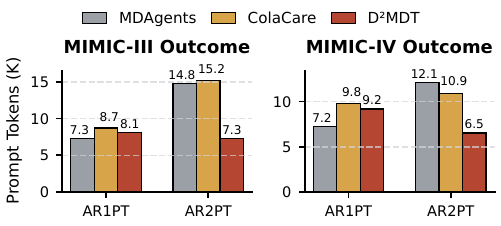}
\caption{The results of efficiency study.}
\label{fig:efficiency}
\vspace{-4mm}
\end{figure}
\begin{figure}[t]
\centering
\includegraphics[width=1\linewidth]{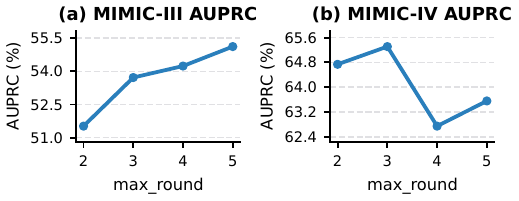}
\caption{Sensitivity analysis of \methodname with respect to \texttt{max\_round}, measured by AUPRC on two datasets.}
\label{fig:max_round_sensitivity}
\vspace{-4mm}
\end{figure}
We show the efficiency comparison of \methodname and competing MDT-based baselines in Figure~\ref{fig:efficiency}, using AR1PT and AR2PT defined in Evaluation Metrics. 
\methodname achieves the best overall efficiency on both MIMIC-III Outcome and MIMIC-IV Outcome, especially in the second-round consultation, which demonstrates the effectiveness of our Residual Deliberation. Although \methodname keeps competitive prompt cost in the first round, its main advantage lies in reducing prompt burden in later rounds.

Compared with other MDT-based baselines, MDAgents and ColaCare usually require more prompt tokens to continue the consultation process. This is mainly because these methods tend to propagate more redundant discussion context across rounds.
In contrast, \methodname focuses on disagreement-aware communication and preserves only the most decision-relevant information for subsequent consultation. Therefore, \methodname can maintain effective collaboration while significantly reducing unnecessary communication overhead.

\subsection{Ablation Study}
\begin{table}[t]
\centering
\tabcolsep=0.1cm 
\caption{Ablation study results on MIMIC-III.}
\label{tab:ablation-III}
\resizebox{\columnwidth}{!}{
\begin{tabular}{l|ccccc}
\toprule
\multirow{2}{*}{\textbf{Model}}
& \multicolumn{5}{c}{\textbf{MIMIC-III Outcome}} \\
\cline{2-6}
& AUPRC ($\uparrow$)
& AUROC ($\uparrow$)
& min(+P, Se) ($\uparrow$)
& AR1PT
& AR2PT \\
\midrule
\textit{w/o} HEM
& 51.87$\pm$4.55 & 84.25$\pm$1.94 & 50.39$\pm$3.84 & 11271.37 & 8925.69 \\
\textit{w/o} RD 
& 53.36$\pm$4.31 & 84.43$\pm$1.94 & 53.36$\pm$3.79 & 8523.41 & 11874.77 \\

\textit{w/o} DR
& 51.77$\pm$3.73 & 83.22$\pm$1.92 & 52.32$\pm$3.11 & 8559.63 & 7200.46 \\
\textit{w/o} DA
& 52.51$\pm$4.88 & 84.87$\pm$1.75 & 53.70$\pm$3.56 & 8645.76 & 7608.78 \\
\midrule
\rowcolor{orange!10}\textbf{\methodname} 
& \textbf{53.72$\pm$4.67} & \textbf{84.56$\pm$2.07} & \textbf{54.68$\pm$3.70}  &\textbf{8145.64} & \textbf{7344.54} \\
\bottomrule
\end{tabular}
}
\vspace{-2mm}
\end{table}

\begin{table}[t]
\centering
\tabcolsep=0.1cm 
\caption{Ablation study results on MIMIC-IV.}
\label{tab:ablation-IV}
\resizebox{\columnwidth}{!}{
\begin{tabular}{l|ccccc}
\toprule
\multirow{2}{*}{\textbf{Model}}
& \multicolumn{5}{c}{\textbf{MIMIC-IV Outcome}} \\
\cline{2-6}
& AUPRC ($\uparrow$) & AUROC ($\uparrow$) & min(+P, Se) ($\uparrow$) & AR1PT & AR2PT \\
\midrule
\textit{w/o} HEM
& 57.93$\pm$5.32 & 86.44$\pm$2.64& 60.91$\pm$4.78 & 9698.24 & 7404.29 \\
\textit{w/o} RD
& 65.15$\pm$5.04& 87.68$\pm$2.01 & 61.44$\pm$4.15 & 9028.31 & 9178.33 \\
\textit{w/o} DR
& 62.99$\pm$4.72 & 87.70$\pm$2.20 & 61.16$\pm$3.98 & 9328.46 & 7846.84 \\
\textit{w/o} DA
& 63.34$\pm$5.11 & 87.47$\pm$2.38 & 60.68$\pm$4.58 & 9704.42 & 7901.63 \\
\midrule
\rowcolor{orange!10}\textbf{\methodname}
& \textbf{65.31$\pm$4.41} & \textbf{87.39$\pm$1.97} & \textbf{63.66$\pm$3.91}& \textbf{9164.02} & \textbf{6534.88} \\
\bottomrule
\end{tabular}
}
\vspace{-3mm}
\end{table}


To investigate whether each designed component contributes to \methodname, we evaluate several ablated variants in Tables~\ref{tab:ablation-III} and~\ref{tab:ablation-IV}.
\begin{itemize}[itemsep=0pt, topsep=1pt, leftmargin=*]
    \item \textbf{\textit{w/o} HEM} removes heterogeneous expert modeling.
    The performance drops clearly, especially in AUPRC and min(+P, Se), showing that heterogeneous experts help capture different aspects of patient status for outcome prediction.

    \item \textbf{\textit{w/o} RD} removes residual deliberation.
    This leads to higher second-round prompt cost and weaker overall results, indicating that residual information helps preserve useful evidence while reducing redundant discussion.

    \item \textbf{\textit{w/o} DR} removes department-aware evidence retrieval.
    This variant uses less targeted evidence. Its performance decrease shows that department-specific retrieval helps each doctor ground the discussion in more relevant clinical context. 

    \item \textbf{\textit{w/o} DA} removes department assignment.
    It verifies that explicit department specialization improves collaboration quality.
\end{itemize}

\subsection{Hyper-parameter Analysis}

\begin{figure*}[!t]
\centering
\includegraphics[width=0.9\linewidth]{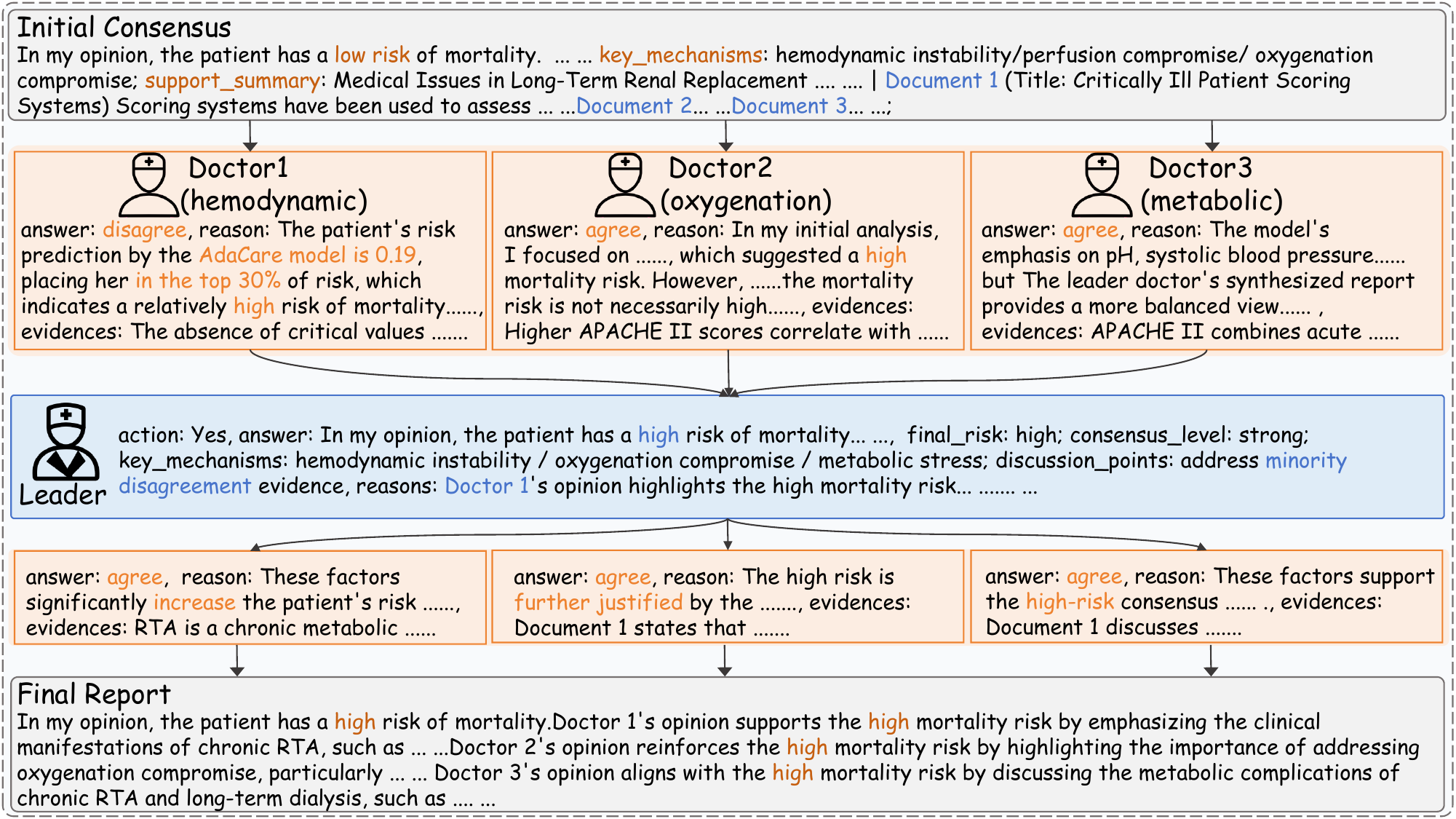}
\caption{Representative \methodname case study. The orange and blue modules denote Doctor Agents and the Leader Agent, respectively, while gray boxes show the initial and final reports during the consultation process.}
\label{fig:case-study}
\vspace{-4mm}
\end{figure*}

We vary the maximum number of consultation rounds (\ie \texttt{max\_round}), and report the corresponding results in Figure~\ref{fig:max_round_sensitivity}. Since AUPRC is the primary metric for imbalanced clinical outcome prediction, we present only the AUPRC trends in the main text for clarity, while the results of other metrics are deferred to Appendix~\ref{sec:appendix-sensitivity}.

As shown in Figure~\ref{fig:max_round_sensitivity}, the two datasets exhibit different trends. On MIMIC-III Outcome, the AUPRC increases steadily as \texttt{max\_round} grows, showing that additional consultation rounds help the doctor agents refine their judgments and reach better consensus. In contrast, 
On MIMIC-IV Outcome, the best AUPRC is achieved at \texttt{max\_round}=3, and further increasing the number of rounds leads to performance degradation. 
This result suggests that most useful collaborative information has already been exchanged in the early rounds, while excessive discussion may introduce redundant context and reduce decision quality.

Considering both predictive performance and the resource cost of additional consultation rounds, we set \texttt{max\_round}=3 for \methodname.

\subsection{Case Study}

To evaluate whether \methodname produces reliable and interpretable consultation traces, we analyze a representative MIMIC-III test patient with cross-system abnormalities. 
The three doctor agents are assigned to \emph{critical care/hemodynamic}, \emph{respiratory/oxygenation}, and \emph{metabolic/general} perspectives, respectively, so that the same EHR trajectory is examined through different clinical roles. 

As illustrated in Figure~\ref{fig:case-study}, initial Consensus indicates the patient has a low mortality risk. In round1, each doctor revisits initial consensus and their own state to express agreement or disagreement. Then leader keeps the disagreement as a compact residual package rather than replaying the full dialogue. In round 2, each doctor revisits only the unresolved minority evidence, and the team finally converges with \texttt{minority\_view: none}. This example highlights two properties of \methodname. First, the decision path remains interpretable because each opinion is explicitly grounded in a department role and concrete physiological evidence. Second, the residual deliberation mechanism improves efficiency by shrinking the follow-up prompt tokens. The full multi-panel qualitative is placed in Appendix~\ref{sec:appendix-case-study}.

\section{Related Works}

\noindent\textbf{Retrieval- and LLM-enhanced Clinical Prediction.}
Recent studies incorporate medical knowledge and language models into EHR prediction. RAM-EHR~\cite{xu2024ram} uses retrieved medical knowledge, CPLLM~\cite{ben2024cpllm} adapts LLMs for disease and readmission prediction, and LLM-based EHR encoders~\cite{hegselmann2025large} serialize structured records into natural language. Clinical LLM systems such as EMERGE~\cite{zhu2024emerge} and MedGemma~\cite{sellergren2025medgemma} further show the potential of LLMs for decision support. Nevertheless, most methods use LLMs as a unified reasoning or encoding module, rather than maintaining parallel specialty-specific assessments.

\vspace{1mm}
\noindent\textbf{MDT-inspired and Agentic Clinical Reasoning.}
LLM-based clinical agents emulate collaborative decision-making through role specialization, debate, and consensus. Systems such as MDAgents~\cite{2024MDAgents}, MAC~\cite{chen2025enhancing}, MAM~\cite{zhou2025mam}, MDTeamGPT~\cite{2025MDTeamGPT}, and ColaCare~\cite{Wang2025ColaCare} demonstrate the value of complementary clinical opinions. However, their predefined roles are often weakly grounded in patient-specific structured EHR evidence, and multi-round discussions may revisit resolved information. \methodname addresses these limitations with department-aware evidence organization and residual deliberation.
\section{Conclusion}
In this paper, we propose \methodname, a Department-aware MultiDisciplinary Team Consultation with Deliberation for Efficient Clinical Prediction. 
Extensive experiments on MIMIC-III and MIMIC-IV demonstrate that \methodname achieves competitive predictive performance while reducing prompt token consumption.
In the future, we will explore stronger medical foundation models, more accurate clinical role modeling, and more adaptive consultation strategies for complex healthcare scenarios.

\section*{Limitations}

This work has several limitations. First, our evaluation is conducted on two ICU outcome prediction benchmarks, MIMIC-III and MIMIC-IV. Although these datasets are widely used for EHR modeling, the current results are still limited to retrospective in-hospital mortality prediction. The effectiveness of \methodname on other clinical tasks, patient populations, and real-world deployment scenarios remains to be further studied.

Second, \methodname relies on LLM-based agents and retrieved medical knowledge to conduct department-aware consultation. Therefore, its reasoning quality can still be affected by the capability of the backbone LLM, the coverage of the external knowledge corpus, and the accuracy of department routing. In addition, although residual deliberation reduces redundant discussion, it may omit useful context when the disagreement state is overly compressed. Future work will explore more adaptive routing and residual construction strategies, together with clinician evaluation in realistic clinical workflows.

Third, as a clinical prediction framework, \methodname may pose potential risks if used beyond research settings. Incorrect predictions or overconfident agent-generated rationales could mislead downstream clinical decision making, especially if the system is treated as a substitute for clinician judgment. In addition, biases inherited from retrospective ICU datasets may lead to uneven performance across patient subgroups, and LLM-based agents may generate incomplete or unsupported reasoning when the retrieved evidence is insufficient. Although our experiments are conducted on de-identified public benchmarks, any real world deployment would require strict privacy protection, external validation, calibration, subgroup fairness evaluation, and continuous clinician oversight. We therefore position \methodname as a decision-support research framework rather than an autonomous diagnostic or treatment system.

\section*{Ethical Considerations}

This work uses MIMIC-III and MIMIC-IV, which are publicly available de-identified clinical datasets distributed through PhysioNet under controlled access. We use these datasets only for research on ICU outcome prediction and follow their data use requirements. We do not attempt to re-identify patients, redistribute the data, or use the data for clinical deployment or patient-level intervention. The proposed framework is intended as a research prototype for clinical decision support rather than an autonomous diagnostic or treatment system.


\bibliography{main}

\clearpage
\appendix
\appendix

\section{Additional Method Details}
\label{sec:appendix-method-details}

\subsection{Semantic Evidence Construction}
\label{sec:appendix-semantic}
For each expert, \methodname computes feature attribution scores on the latest-visit structured EHR input and ranks variables by their absolute contribution to the expert risk estimate. In the current implementation, these attributions are produced by SHAP and saved as feature-level importance weights. \methodname keeps the top important variables for each expert and verbalizes each retained variable with two observable signals from the raw EHR sequence: its latest value and a short description of the local temporal trend near the latest observation time (e.g., persistent elevation, recent drop, abnormal fluctuation). This step converts model-specific numerical evidence into a compact textual card that can be interpreted by doctor agents without exposing latent vectors directly.
Table~\ref{tab:appendix_structured_fields} summarizes the retained fields.

\begin{table}[H]
\centering
\caption{Fields retained in \methodname's structured semantic and residual summaries.}
\label{tab:appendix_structured_fields}
\small
\begin{tabular}{p{0.26\columnwidth}p{0.66\columnwidth}}
\toprule
\textbf{Component} & \textbf{Retained fields} \\
\midrule
Evidence card & latest important variables, recent temporal trends, and expert-specific risk cues derived from the EHR encoder and retrieved knowledge \\
Compact doctor card & \texttt{risk\_vote}, \texttt{top\_mechanisms}, \texttt{support\_features}, \texttt{counter\_evidence}, \texttt{confidence}, \texttt{need\_escalation} \\
Routing summary & cross-expert abnormal findings, merged salient variables, and whether doctor risk votes are aligned or conflicting \\
Compact leader report & \texttt{final\_risk}, \texttt{consensus\_level}, \texttt{key\_mechanisms}, \texttt{support\_summary}, \texttt{minority\_view}, \texttt{discussion\_points} \\
\bottomrule
\end{tabular}
\end{table}
\subsection{Depatment routing}
\label{sec:appendix-department}
Department assignment is implemented as a lightweight rule-based scorer. For each patient, \methodname extracts three routing sources. The first is the abnormal numerical set from \texttt{raw\_x}. Concretely, for each continuous variable listed in Table~\ref{tab:DeptMDT_variables}, \methodname compares the latest observed value against that patient's own recent trajectory by computing a deviation score relative to the within-patient median and standard deviation; the variable is marked abnormal when this normalized deviation exceeds a preset threshold. The second source is the important-feature set. \methodname reads the stored feature-importance file, ranks variables by the absolute SHAP-based importance weight, and keeps the top-$k$ variables for routing. The third source is the context-risk signal. This signal is not produced by a text encoder; instead, the code applies rule-based cue extraction to the patient context text and parses coarse indicators such as predicted mortality risk, whether the patient is in ICU, and whether the case falls into the top 30\% risk range.

For a department candidate $d$, the routing score is the weighted sum of three parts: abnormal-numeric score, important-feature score, and context-risk score. The first two are accumulated through a feature-to-department match function $\rho_d(f)$. In the current implementation, $\rho_d(f)$ is computed by exact feature-to-department mapping when the variable name is one of the predefined \methodname variables in Table~\ref{tab:DeptMDT_variables}, and otherwise by keyword-overlap matching between the normalized variable name and the department keyword lexicon. The third term is a department-specific bonus $c_d$ triggered by the parsed risk cues. In the current implementation, the three weights are 1.4, 0.9, and 0.7, respectively. The system computes scores for all candidate departments, but the dynamic assignment step only ranks the non-fixed departments because the first doctor is always anchored to the critical-care/hemodynamic role.
\begin{table*}[t]
\centering
\renewcommand{\arraystretch}{1.1}
\setlength{\tabcolsep}{6pt}
\begin{tabular}{|l|l|l|}
\hline
\rowcolor{orange!30}
\textbf{Variable} & \textbf{Impute Value} & \textbf{Modeled as} \\
\hline
Capillary refill rate & -1 & categorical \\
\hline
Glascow coma scale eye opening & -1 & categorical \\
\hline
Glascow coma scale motor response & -1 & categorical \\
\hline
Glascow coma scale total & -1 & categorical \\
\hline
Glascow coma scale verbal response & -1 & categorical \\
\hline
Diastolic blood pressure & z-score & continuous \\
\hline
Fraction inspired oxygen & z-score & continuous \\
\hline
Glucose & z-score & continuous \\
\hline
Heart Rate & z-score & continuous \\
\hline
Height & z-score & continuous \\
\hline
Mean blood pressure & z-score & continuous \\
\hline
Oxygen saturation & z-score & continuous \\
\hline
Respiratory rate & z-score & continuous \\
\hline
Systolic blood pressure & z-score & continuous \\
\hline
Temperature & z-score & continuous \\
\hline
Weight & z-score & continuous \\
\hline
pH & z-score & continuous \\
\hline
\end{tabular}
\caption{The 17 selected clinical variables. The second column lists the imputation values used in preprocessing, and the third column describes how variables are modeled (categorical or continuous).}
\label{tab:DeptMDT_variables}
\end{table*}

\subsection{Prompt Template}
\label{sec:appendix-prompt-template}
\methodname uses a multi-stage prompt pipeline adapted to department-aware consultation and residual deliberation. We report example prompt templates below (placeholders in \texttt{<>}). The stages align with the initial consensus and later-round residual updates described in Section~2.

\begin{figure*}[t]
\centering
\begingroup
\setlength{\fboxsep}{7pt}
\newcommand{\methodnamepromptbox}[2]{%
\fcolorbox{black!20}{gray!10}{%
\begin{minipage}[t]{0.455\textwidth}
\vspace{0pt}
\colorbox{black!15}{\parbox{\dimexpr\linewidth-2\fboxsep\relax}{\raggedright\footnotesize\bfseries #1}}\par\medskip
\raggedright\footnotesize #2
\vspace{1mm}
\end{minipage}}}


\methodnamepromptbox{Initial doctor prompt.}{%
\textbf{System.} An experienced doctor analyzes multivariate EHR, model-predicted mortality risk, feature importance, and retrieved medical knowledge. The patient record is prefixed with an assigned department hint.\par\medskip
\textbf{Inputs.} Retrieved knowledge \texttt{context}; patient record and model analysis \texttt{hcontext}.\par\medskip
\textbf{Output JSON.} Fields \texttt{logit}, \texttt{analysis}, and \texttt{evidences}.\par\medskip
\textbf{Derived compact doctor card.} \methodname builds a structured card with \texttt{risk\_vote}, \texttt{top\_mechanisms}, \texttt{support\_features}, \texttt{counter\_evidence}, \texttt{confidence}, and \texttt{need\_escalation}.}
\hfill
\methodnamepromptbox{Initial leader synthesis.}{%
\textbf{System.} The leader reads all doctor opinions, checks whether they are clinically reasonable, and writes a synthesized patient-level report.\par\medskip
\textbf{Inputs.} Patient basic information \texttt{patient\_info}; aggregated doctor statements generated by \texttt{generate\_doctors\_prompt(..., is\_initial=True)}.\par\medskip
\textbf{Output JSON.} Fields \texttt{answer}, \texttt{report}, and \texttt{evidences}.\par\medskip
\textbf{Derived compact leader report.} The report is compressed into \texttt{final\_risk}, \texttt{consensus\_level}, \texttt{key\_mechanisms}, \texttt{support\_summary}, \texttt{minority\_view}, and \texttt{discussion\_points}.}

\par\vspace{2mm}

\methodnamepromptbox{Later-round doctor revision.}{%
\textbf{System.} Each doctor revisits the leader conclusion and provides an updated view.\par\medskip
\textbf{Default inputs.} Retrieved knowledge \texttt{context}; the doctor's previous analysis \texttt{analysis}; leader opinion \texttt{opinion}; leader report \texttt{report}.\par\medskip
\textbf{Delta-discussion inputs.} For rounds \(r\ge 2\), \methodname substitutes these slots with compact residual states: doctor compact text, disagreement points, and compact leader report.\par\medskip
\textbf{Output JSON.} Fields \texttt{answer} (agree/disagree), \texttt{confidence} (1--3), \texttt{reason}, and \texttt{evidences}.}
\hfill
\methodnamepromptbox{Later-round leader update and stopping decision.}{%
\textbf{Implemented summary prompt.} Read the previous synthesized report plus all revised doctor statements, decides whether another round is needed, and updates the patient-level report.\par\medskip
\textbf{Inputs.} Previous leader report \texttt{latest\_info}; revised doctor statements from \texttt{generate\_doctors\_prompt(..., is\_initial=False)}.\par\medskip
\textbf{Output JSON.} Fields \texttt{action}, \texttt{answer}, \texttt{report}, and \texttt{reasons}.\par\medskip
\textbf{Auxiliary stopping prompt.} A lighter stop/continue prompt output contains only \texttt{action} and \texttt{reason}.}
\endgroup
\caption{Implementation-faithful prompt and compact-state templates used in \methodname. The first row shows the actual first-round doctor and leader prompts; the second row shows the later-round revision and stopping/update stages, including the compact semantic states injected by residual deliberation.}
\label{fig:appendix-prompt-template}
\end{figure*}

\subsection{Inference and Training}
\label{sec:appendix:method_algorithm}

The inference pipeline of \methodname consists of four stages. 
First, three doctor agents independently analyze the same patient using different EHR expert backbones and their assigned departments. 
Second, the leader agent summarizes these initial opinions into a synthesized report and produces an initial risk judgment. 
Third, the doctor agents enter iterative consultation rounds, where they reconsider their opinions according to the leader report and the current disagreement points. 
Finally, the leader agent revises the final logits based on the completed consultation outputs.

For prediction training, the collaboration outputs are converted into fusion features and used to train a lightweight fusion model. The fusion model takes the EHR embeddings and text embeddings produced by the collaboration stage as input, and learns the final prediction under the same train/validation/test split protocol.

\begin{algorithm}[t]
\caption{Training and Inference of D$^2$MDT}
\label{alg:methodname_train_infer}
\raggedright
\begin{algorithmic}[1]
\Require Processed EHR dataset $\mathcal{D}$, specialist encoder set $\mathcal{M}=\{\text{AdaCare}, \text{MCGRU}, \text{RETAIN}\}$, LLM-based collaboration module $\mathcal{C}$
\Ensure Final prediction score $\hat{y}$ for each patient

\Statex \textbf{Stage I: Train specialist EHR encoders}
\For{each encoder $m \in \mathcal{M}$}
    \State Train $m$ on the training split of $\mathcal{D}$
    \State Select the best checkpoint according to validation AUPRC
    \State Use the best checkpoint to extract patient embeddings on validation and test splits
\EndFor

\Statex \textbf{Stage II: Construct fusion data}
\For{each patient $p$}
    \State Collect specialist EHR embeddings $\{ \mathbf{e}_p^{(m)} \}_{m \in \mathcal{M}}$
    \State Generate the final collaborative report $r_p$ using $\mathcal{C}$
    \State Encode $r_p$ into text embedding $\mathbf{t}_p$
    \State Build fusion sample $(\{\mathbf{e}_p^{(m)}\}, \mathbf{t}_p, y_p)$
\EndFor
\State Split fusion data into train/validation/test sets
\State Train the fusion predictor with binary cross-entropy loss
\State Select the best fusion checkpoint according to validation AUPRC

\Statex \textbf{Inference}
\For{each test patient $p$}
    \State Obtain specialist embeddings $\{ \mathbf{e}_p^{(m)} \}_{m \in \mathcal{M}}$
    \State Generate collaborative report $r_p$ and encode it as $\mathbf{t}_p$
    \State Fuse $\{\mathbf{e}_p^{(m)}\}$ and $\mathbf{t}_p$ to obtain prediction $\hat{y}_p$
\EndFor
\end{algorithmic}
\end{algorithm}

\section{Additional Experimental Details}
\label{sec:appendix:exp_details}
\subsection{Datasets and Splits}
\label{sec:appendix:datasets}
We evaluate \methodname on two widely used ICU outcome prediction benchmarks, \textbf{MIMIC-III Outcome}~\cite{johnson2016mimic} and \textbf{MIMIC-IV Outcome}~\cite{johnson2023mimic}. Both datasets are derived from real-world intensive care unit electronic health records (EHRs), and the task is in-hospital outcome prediction based on structured multivariate clinical time series. The input records contain longitudinal patient measurements together with basic clinical attributes, enabling the model to reason over temporal patient states under realistic clinical settings. The statistics of the dataset splits are summarized in Table~\ref{tab:dataset_split}.

\begin{table}[t]
\centering
\caption{Statistics of the experimented datasets after preprocessing. The \# Samples column reports the number of patient records and their percentage in each split.}
\label{tab:dataset_split}
\begin{tabular}{ccc}
\toprule
\textbf{Dataset} & \textbf{Split} & \textbf{\# Samples} \\
\midrule
\multirow{3}{*}{MIMIC-III} 
& Train & 19,026 (90.00\%) \\
& Val   & 1,057 (5.00\%) \\
& Test  & 1,057 (5.00\%) \\
\midrule
\multirow{3}{*}{MIMIC-IV} 
& Train & 13,302 (90.00\%) \\
& Val   & 739 (5.00\%) \\
& Test  & 739 (5.00\%) \\
\bottomrule
\end{tabular}
\end{table}

To enhance clinical reasoning, we additionally introduce the \textbf{Merck Manual of Diagnosis and Therapy (MSD)}~\cite{porter2011merck} as an external medical guideline corpus. During inference, \methodname retrieves relevant MSD knowledge as supportive evidence for doctor agents, so that the multi-agent discussion is grounded not only in patient-specific EHR signals but also in explicit medical knowledge. This external corpus is used only at inference time and does not alter the benchmark definition or the training labels.

\subsection{Baseline categories}
\label{sec:appendix:baseline}
We organize the compared methods into three groups according to their modeling assumptions and clinical decision-making mechanisms.

\vspace{1mm}
\noindent\textbf{Deep learning models.}
This group contains strong structured EHR prediction methods that directly model multivariate clinical time series without explicit LLM-based collaboration.

\begin{itemize}[leftmargin=*, topsep=2pt, itemsep=1pt, parsep=0pt]
    \item \textbf{AdaCare}~\cite{2020AdaCare} is an interpretable clinical time-series model based on scale-adaptive feature extraction and recalibration. It captures patient status dynamics while highlighting clinically important variables, and is therefore a representative deep learning baseline for EHR outcome prediction.
    \item \textbf{RETAIN}~\cite{2016RETAIN} is a reverse-time attention model for healthcare prediction. It is widely adopted as a strong interpretable EHR baseline because it assigns visit-level and variable-level attention weights while preserving competitive predictive performance.
    \item \textbf{PAI}~\cite{2024Learnable} is a recent EHR prediction framework that replaces explicit missing-value imputation with learnable prompts as pseudo-imputation. Since ICU EHR data contain substantial missingness, PAI serves as a strong recent baseline for evaluating whether \methodname remains effective under modern missing-data-aware modeling.
\end{itemize}

\noindent\textbf{LLM-based models.}
This group introduces external medical knowledge or large language model reasoning into the prediction process, but does not explicitly implement a department-aware MDT collaboration mechanism as \methodname does.

\begin{itemize}[leftmargin=*, topsep=2pt, itemsep=1pt, parsep=0pt]
    \item \textbf{EMERGE}~\cite{zhu2024emerge} is a retrieval-augmented multimodal EHR framework that extracts medical entities from time-series data and clinical notes, aligns them with external medical knowledge, and generates task-relevant patient summaries for downstream prediction. We include EMERGE because it represents a strong RAG-style LLM-enhanced EHR prediction pipeline.
    \item \textbf{MedGemma Zero-Shot} and \textbf{MedGemma Few-Shot}~\cite{sellergren2025medgemma} use the MedGemma medical foundation model directly for clinical reasoning under zero-shot and few-shot prompting settings, respectively. These baselines are used to test whether general-purpose medical foundation models can be directly transferred to structured ICU outcome prediction without additional task-specific collaboration design.
\end{itemize}

\noindent\textbf{MDT-based models.}
This group is the closest comparison family to \methodname, since all methods explicitly simulate multi-doctor consultation or multi-agent medical decision-making.

\begin{itemize}[leftmargin=*, topsep=2pt, itemsep=1pt, parsep=0pt]
    \item \textbf{MDAgents}~\cite{2024MDAgents} is an adaptive medical multi-agent framework that dynamically assigns collaboration structures among LLM agents for medical decision-making. It serves as a representative general-purpose medical multi-agent baseline.
    \item \textbf{MDTeamGPT}~\cite{2025MDTeamGPT} is a self-evolving LLM-based MDT consultation framework with consensus aggregation, residual discussion structure, and experience accumulation. We compare against MDTeamGPT because it emphasizes multi-round consultation efficiency and knowledge reuse, which are closely related to the design goals of \methodname.
    \item \textbf{ColaCare}~\cite{Wang2025ColaCare} is the most direct baseline to \methodname. It integrates EHR expert models, LLM doctor agents, a meta-agent, and MSD-based retrieval within an MDT-style collaborative consultation framework. Compared with ColaCare, \methodname further introduces patient-specific department assignment and compact disagreement-focused discussion, so this comparison directly evaluates the contribution of our design.
\end{itemize}

Whenever a baseline is reproduced locally, we use the same dataset split, evaluation pipeline, and prediction post-processing as \methodname. This controlled setup keeps the comparison focused on modeling and collaboration differences rather than implementation-side evaluation discrepancies.
\subsection{Metrics details}
\label{sec:appendix:metrics}

AUROC~\cite{mcdermott2024closer}: This metric is our primary consideration in binary
classification tasks due to its widespread use in clinical settings
and its effectiveness in handling imbalanced datasets.
AUPRC~\cite{kim2022empirical}: The AUPRC is particularly useful for evaluating perfor-
mance in datasets with a significant imbalance between classes.
min(+P, Se)~\cite{ma2022patient}: This composite metric represents the minimum
value between precision (+P) and sensitivity (Se), providing a
balanced measure of model performance.
All these three metrics are the higher the better.

\textbf{AR1PT} and\textbf{AR2PT} as the average prompt tokens consumed in round 1 and round 2, respectively. AR2PT being smaller than AR1PT indicates that residual discussion is effective.

\subsection{Implementation stack.}
\label{sec:appendix:implementation}
\methodname uses three doctor agents and one leader agent. The three doctor agents are paired with AdaCare, MCGRU, and RETAIN, respectively, while the leader agent summarizes specialist opinions and produces the final consensus report. In the full \methodname setting, the consultation pipeline enables department-aware prompting, department-specific retrieval, structured doctor cards, semantic doctor cards, compact leader reports, and residual disagreement discussion. All consultation agents are driven by \textbf{Qwen3-8B}. For retrieval-augmented consultation, we use \textbf{MedCPT} together with the \textbf{MSD} corpus.

For multimodal prediction, the final leader report is encoded by \textbf{GatorTron-base}. Each report is tokenized with a maximum length of 512 and represented by the last-layer \texttt{[CLS]} embedding, yielding a 1024-dimensional text feature. This text feature is concatenated with the three 128-dimensional EHR embeddings from AdaCare, MCGRU, and RETAIN. The fusion module is a lightweight MLP that first projects the concatenated multimodal vector to a 128-dimensional hidden representation, applies a GELU activation, and then outputs the final risk probability through a linear layer followed by a sigmoid.

We train the fusion model with AdamW using a learning rate of $1\times10^{-3}$ and binary cross-entropy loss. The maximum number of training epochs is 50, the early-stopping patience is 10, and model selection is based on validation AUPRC. The batch size is 128. We fix the random seed to 42 for reproducibility. Unless otherwise specified, the maximum number of consultation rounds is set to 3 for both MIMIC-III and MIMIC-IV.

\subsection{Hardware and software environment.}
The current local environment provides multiple \textbf{NVIDIA GeForce RTX 4090 24GB GPUs} (driver version 590.48.01). The repository requirements specify torch 2.3.1, lightning 2.3.3, transformers 4.42.4, numpy 1.26.4, pandas 2.2.2, scikit\_learn 1.5.1, shap 0.46.0, faiss\_cpu 1.8.0.post1, and faiss\_gpu 1.7.2. In practice, the exact Python runtime and serving backend may vary across machines, but all reported experiments follow the same repository-level configuration, retrieval stack, and evaluation scripts.





\subsection{Sensitivity Analysis}
\label{sec:appendix-sensitivity}
\begin{figure}[!t]
\centering
\includegraphics[width=1\linewidth]{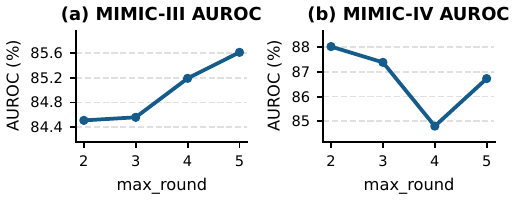}
\caption{Sensitivity analysis of AUROC under different values of \texttt{max\_round}.}
\label{fig:auroc}
\vspace{-5mm}
\end{figure}

\begin{figure}[!t]
\centering
\includegraphics[width=1\linewidth]{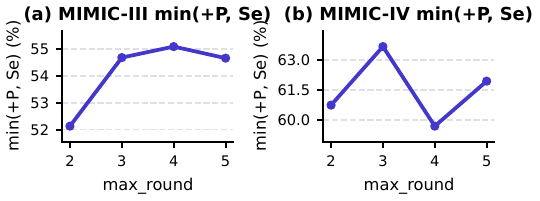}
\caption{Sensitivity analysis of min(+P, Se) under different values of \texttt{max\_round}.}
\label{fig:minps}
\vspace{-5mm}
\end{figure}

\begin{figure}[!t]
\centering
\includegraphics[width=1\linewidth]{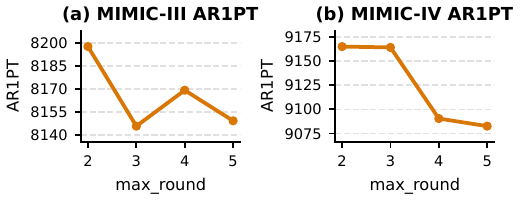}
\caption{Sensitivity analysis of communication cost AR1PT under different values of \texttt{max\_round}.}
\label{fig:ar1pt}
\vspace{-5mm}
\end{figure}

\begin{figure}[!t]
\centering
\includegraphics[width=1\linewidth]{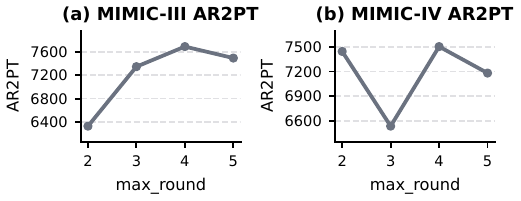}
\caption{Sensitivity analysis of predictive metrics under different values of min(+P, Se).}
\label{fig:ar2pt}
\vspace{-5mm}
\end{figure}

Figure~\ref{fig:auroc}, Figure~\ref{fig:minps}, Figure~\ref{fig:ar1pt} and Figure~\ref{fig:ar2pt} report the remaining sensitivity results that are not shown in the main text. These figures complement the main-text AUPRC analysis by showing how predictive robustness and communication cost change as the maximum consultation round varies.


\subsection{Case Study}
\label{sec:appendix-case-study}
Case study of a patient, illustrating how the discussion evolves from the initial consensus to the final report. Based on the initial consensus, the leader-level synthesis initially assessed the patient as low risk, emphasizing hemodynamic instability, perfusion compromise, and oxygenation compromise. In the subsequent round 1 review, however, Doctor 1 disagreed with this low-risk conclusion, arguing that the AdaCare score of 0.19, the patient’s top-30\% risk ranking among ICU patients, and the possibility of physiological instability associated with long-term renal replacement therapy, metabolic derangements, and acid-base disturbance together suggested a higher mortality risk. Doctors 2 and 3, in contrast, agreed with the low-risk synthesis, considering the observed abnormalities insufficient on their own to justify a high-risk interpretation.
After aggregating these responses, the leader produced the round1 summary, which served as the effective initial consensus for the next stage. At this point, the leader revised the conclusion to high risk, identified hemodynamic instability, oxygenation compromise, and metabolic stress as the key mechanisms, and explicitly marked the remaining minority disagreement as the focus for further discussion. Because this unresolved disagreement persisted, the case proceeded to round 2.
In round 2, the system forwarded only compact structured summaries and the residual disagreement point, rather than replaying the full first-round discussion. Under this focused setting, all three doctors converged on a high-risk judgment. Doctor 1 emphasized chronic renal tubular acidosis and life-threatening electrolyte disturbances, Doctor 2 highlighted oxygenation compromise and respiratory failure, and Doctor 3 stressed the metabolic complications of chronic RTA and long-term dialysis. The final leader report therefore concluded that the patient had a high risk of mortality with strong consensus and no remaining minority view, showing that the second-round residual discussion helped resolve the initial disagreement and refine the final clinical assessment.

\end{document}